\documentstyle[11pt,aaspp4]{article}
\slugcomment{Submitted to Astron. J.}
\lefthead{Haisch et al.}
\righthead{An Infrared Multiplicity Survey of Class I/Flat-Spectrum Systems in the $\rho$ Ophiuchi and Serpens Molecular Clouds}

\begin{document}

\title{An Infrared Multiplicity Survey of Class I/Flat-Spectrum Systems in the $\rho$ Ophiuchi and Serpens Molecular Clouds}

\author{Karl E. Haisch Jr.\altaffilmark{1,2}}
\affil{NASA Ames Research Center, Mail Stop 245-6, Moffett Field, California  94035-1000, khaisch@mail.arc.nasa.gov}

\and

\author{Mary Barsony\altaffilmark{3,4,5}}
\affil{Jet Propulsion Laboratory, Mail Stop 169-327, 4800 Oak Grove Drive, Pasadena, California  91109, fun@uhuru.jpl.nasa.gov}

\and

\author{Thomas P. Greene\altaffilmark{2}}
\affil{NASA Ames Research Center, Mail Stop 245-6, Moffett Field, California  94035-1000, tgreene@mail.arc.nasa.gov}

\and

\author{Michael E. Ressler\altaffilmark{3,4}}
\affil{Jet Propulsion Laboratory, Mail Stop 169-327, 4800 Oak Grove Drive, Pasadena, California  91109, Michael.E.Ressler@jpl.nasa.gov}

\altaffiltext{1}{National Research Council Resident Research Associate}

\altaffiltext{2}{Visiting Astronomer at the Infrared Telescope Facility which is
operated by the University of Hawaii under contract to the National Aeronautics
and Space Administration.}

\altaffiltext{3}{Visiting Astronomer at the W. M. Keck Observatory, which is operated as a scientific partnership among the California Institute of Technology, the University of California, and the National Aeronautics and Space Administration. The Observatory was made possible by the generous financial support of the W. M. Keck Foundation.}

\altaffiltext{4}{Observations with the Palomar 5 m telescope were obtained under a collaborative agreement between Palomar Observatory and the Jet Propulsion Laboratory.}

\altaffiltext{5}{Space Science Institute, 3100 Marine Street, Suite A353, Boulder, CO 80303-1058}

\begin{abstract}

We present new near- and mid-IR observations of 19 Class I/flat-spectrum young stellar objects in the nearby $\rho$ Ophiuchi ($d$ = 125 pc) and Serpens ($d$ = 310 pc) dark clouds. These observations are part of a larger systematic infrared multiplicity survey of Class I/flat-spectrum objects in the nearest dark clouds. We find 7/19 (37\% $\pm$ 14\%) of the sources surveyed to be multiple systems over a separation range of $\sim$ 150 -- 1800 AU. This is consistent with the fraction of multiple systems found among older pre-main-sequence stars in each of the Taurus, $\rho$ Ophiuchi, Chamaeleon, Lupus, and Corona Australis star-forming regions over a similar separation range. However, solar-type main-sequence stars in the solar neighborhood have a fraction approximately one-third that of our Class I/flat-spectrum sample (11\% $\pm$ 3\%). This may be attributed to evolutionary effects or environmental differences. An examination of the spectral energy distributions (SEDs) of the SVS 20 and WL 1 binaries reveals that the individual components of each source exhibit the same SED classifications, similar to what one typically finds for binary TTS systems, where the companion of a classical TTS also tends to be of the same SED type.
\end{abstract}


\section{Introduction}

The past two decades have witnessed substantial progress in the study of star formation and its aftermath, the birth of planetary systems. Observations have led the way, with much of the credit due to both spaceborne missions, such as the Infrared Astronomical Satellite (IRAS), the Infrared Space Observatory (ISO), and the Hubble Space Telescope (HST), and to ground-based infrared (IR) imaging surveys from NOAO/SQIID, the NASA Infrared Telescope Facility (IRTF) and other facilities. Advances in detection and instrumentation have been matched by equally exciting theoretical developments. Researchers have delineated key features of a young star's structure and the manner in which it affects its surroundings, both thermally and mechanically.

As we broaden our focus from individual stars, we find that there is a universal tendency for these objects to form in groups, rather than as isolated entities. On the largest scales, IR surveys have revealed stellar aggregates still embedded in their parent molecular clouds (e.g., \cite{zmw93}). These must eventually become optically visible associations or bound clusters. However, the smallest entities created within such groups are generally $not$ single stars. We have known for many years that the majority of solar-to-late-type field stars are binaries or multiples (\cite{al76}; \cite{dm91}; \cite{fm92}), but only in the past decade have significant numbers of pre-main-sequence (PMS) T Tauri stars (TTSs) been surveyed for multiplicity. These recent IR surveys have shown that young low-mass stars have binary fractions that are greater than or equal to that of the field (e.g., \cite{gnm93}; \cite{math94}; \cite{simon95}). These pairs appear to be coeval, that is, at the same evolutionary age (\cite{hss94}; \cite{bz97}). Recent observations at millimeter continuum wavelengths also find multiple sources at the origins of extended molecular outflows and optical jets (\cite{lmw97}). The current findings set this fact in its proper evolutionary context. Most stars $form$ as part of a binary system, rather than find their partners later in life.

The large binary frequency in visible, PMS stars prompts us to investigate even younger systems. Recent sub-arcsecond surveys of young stars have almost exclusively focused on PMS TTSs and have virtually ignored younger systems with Class I/flat-spectrum energy distributions. Thus we know very little about the multiplicity of self-embedded young stars. Is multiplicity the rule even among more deeply embedded objects, those still forming out of their parent clouds? What is their binary frequency, and how does this compare to that of PMS stars? What are the energy distributions of their components? New multiplicity surveys of known embedded low-mass stars must be conducted to address these questions.

We have initiated the first systematic near-to-mid-IR multiplicity survey of $\sim$ 100 Class I/flat-spectrum young stellar objects in six nearby (d $\la$ 300 pc) dark clouds. Our survey has several purposes: 1) to study the binary fractions and separations of protostars; 2) to diagnose the luminosities and evolutionary states of their components; and 3) use our near-IR photometry (in conjunction with existing IRTF and Keck spectra) to place protostars in Hertzsprung-Russell (H-R) diagrams to diagnose their stellar properties.

All of the sources in our survey were selected such that they have either Class I (protostellar) or flat-spectrum (slightly more evolved than Class I) spectral energy distributions (SEDs; see \cite{als87}; hereafter ALS87, 1988; \cite{lada87}). These objects are the direct evolutionary progenitors of the well-studied TTSs, and many are reasonably bright in the near-IR (m$_{K}$ $\sim$ 10). Although these objects have peak fluxes at far-IR wavelengths, there are no far-IR high-resolution observatories (0.85 -- 3.0 m for SIRTF -- SOFIA) currently operating. Also, millimeter wavelength observations of the very few known Class I binaries show that their envelopes can be so large that they overlap, making it difficult to discern the individual components at these long wavelengths (\cite{lmw00}).

In this contribution, we present the initial results of our survey. Specifically, we discuss our near- and mid-IR observations of 19 Class I/flat-spectrum objects in the $\rho$ Ophiuchi ($d$ = 125 pc; \cite{kh98}) and Serpens ($d$ = 310 pc; \cite{del91}) dark clouds. We discuss the observations and data reduction procedures in $\S$2. In $\S$3, we present the results of our survey, and discuss the results in $\S$4. We summarize our primary results in $\S$5.

\section{Observations and Data Reduction}

Two different sets of infrared observations were obtained for this survey. The near-IR $JHKL$-band (1.25, 1.65, 2.2, and 3.5 $\mu$m) observations of each source were obtained with the NSFCAM 256$\times$256 InSb facility array camera (\cite{ray93}; \cite{shu94}) on the NASA IRTF 3 m telescope on Mauna Kea, Hawaii. These data were supplemented with mid-IR $N$-band (10.8 $\mu$m) observations taken with MIRLIN, JPL's 128$\times$128 pixel Si:As camera (\cite{ress94}), on both the Palomar Observatory 5 m and Keck II 10 m telescopes. A summary of all observations is shown in Table~\ref{table1}.

\subsection{NSFCAM-IRTF Near-IR Observations}

The $J, H, K$ and $L$-band observations of each source were made during the period 2001 July 08 - 10. For our study, we used a plate scale of 0\farcs148 pixel$^{-1}$ with a corresponding field of view of approximately 38\arcsec$\times$38\arcsec. The average FWHM for all observations was $\sim$ 1\arcsec.

Each source was observed in a five point dither pattern (a 2 $\times$ 2 square with a point at the center) with 12\arcsec \hspace*{0.05in}offsets between the corners of the square. At each dither position, the telescope was nodded to separate sky positions 30\arcsec \hspace*{0.05in}north of each target observation. Typical total integration times ranged from 10 seconds to 15 minutes for the $JHK$ data, and 60 seconds at $L$-band.

All $JHKL$ data were reduced using the Image Reduction and Analysis Facility (IRAF)\footnote[5]{IRAF is distributed by the National Optical Astronomy Observatories, which are operated by the Association of Universities for Research in Astronomy, Inc., under cooperative agreement with the National Science Foundation.}. The individual sky frames were normalized to produce flat fields for each target frame. All target frames were processed by subtracting the appropriate sky frames and dividing by the flat fields. Finally all target frames were registered and combined to produce the final images of each object.

\subsection{MIRLIN-Keck-Palomar Mid-IR Observations}

All mid-IR $N$-band ($\lambda_0=10.78\ \mu$m, $\Delta \lambda=5.7\ \mu$m) data were obtained as part of a larger, mid-IR imaging survey of the $\rho$ Oph cloud core's embedded population (\cite{brm02}). The plate scale and field of view of MIRLIN at Keck II are 0\farcs138 pixel$^{-1}$ and 17\farcs7 $\times$ 17\farcs7, respectively. Corresponding values at the Palomar 5-meter are 0\farcs15 pixel$^{-1}$ and 19\farcs2 $\times$ 19\farcs2. For reference, the full-width at half-maximum of a diffraction-limited image at $N$-band is $\sim$ 0\farcs25 at Keck II and 0\farcs47 at the Palomar 5-meter.

Data were acquired with traditional mid-IR chopping and nodding techniques. Specifically, the telescope's secondary mirror was chopped 8\arcsec \hspace*{0.05in}in a north-south direction, at a rate of a few Hz; then the entire telescope was nodded 8\arcsec \hspace*{0.05in}east-west in order to remove residual differences in the background level. Total on-source integration times were typically 24 -- 25 seconds at each telescope for the program sources.
On-source integration times consisted of several hundred to a thousand coadded chop pairs, with 5 -- 6 msec integration times per frame. The raw images were background-subtracted, shifted, and co-added with our in-house IDL routine ``MAC'' (match-and-combine).

\subsection{Near-IR Source Photometry and Calibration}

Aperture photometry was performed using the PHOT routine within IRAF. An aperture of 4 pixels in radius was used for all target photometry, and a 10 pixel radius was used for the standard star photometry. Sky values around each source were determined from the mode of intensities in an annulus with inner and outer radii of 10 and 20 pixels respectively. Our choice of aperture size for our target photometry insured that the individual source fluxes were not contaminated by the flux from companion stars, however they are not large enough to include all the flux from a given source. In order to account for this missing flux, aperture corrections were determined using the MKAPFILE routine within IRAF. The instrumental magnitudes for all sources were corrected to account for the missing flux.

Photometric calibration was accomplished using the list of standard stars of Elias et al. (1982). The standards were observed on the same nights and through the same range of airmasses as the target sources. Zero points and extinction coefficients were established for each night. All magnitudes and colors were transformed to the CIT system using Mauna Kea to NSFCAM and NSFCAM to CIT photometric color transformation equations from http://irtf.ifa.hawaii.edu/Facility/nsfcam/mkfilters.html, http://irtf.ifa.hawaii.edu/Facility/nsfcam/hist/color.html and the NSFCAM User's Guide. The photometric accuracy for all observations is typically good to within $\pm$ 0.10, 0.03, 0.03, and 0.06 magnitudes at $J$, $H$, $K$, and $L$-band respectively. GY 224 has a $J$-band magnitude uncertainty of $\pm$ 0.32 mag, and the $H$-band magnitude uncertainty for EC 53 is $\pm$ 0.08.

\subsection{Mid-IR Source Photometry and Calibration}

Flux standards were $\alpha$ Corona Borealis ($N$ = $+$2.19) and $\gamma$ Aquilae ($N$ = --0.78) for the Keck II run, and $\gamma$ Aquilae and $\phi$ Ophiuchi ($N$ = $+$2.20) for the Palomar data. These were also used for airmass monitoring during each night of observing.

Photometry for the standard stars was performed in 14-pixel radius apertures (corresponding to a 1\farcs9 radius software aperture at Keck II and a 2\farcs1 radius aperture at Palomar). A straight-line fit to the instrumental magnitudes as a function of airmass for each night resulted in the determination of the zero-point offsets and airmass corrections. No airmass correction was required for the Keck II data, whereas typical airmass corrections at Palomar were of order 0.1 -- 0.2 mags/airmass. The photometric consistency between all standards during a given night's observing was typically of order $\pm$ 0.04 magnitudes. By adding the errors in the zero-point offsets, the airmass corrections, the aperture corrections, and the uncertainties in the magnitudes of the standards in quadrature and taking the square root, we estimate the total photometric accuracy of the Keck data to be good to $\pm$ 0.06 magnitudes at $N$-band. The Palomar data have a photometric accuracy good to $\pm$ 0.08 magnitudes at $N$-band. To convert these errors to Janskys, note that 0.00 magnitudes at $N$-band with MIRLIN corresponds to 33.4 Jy. Photometry for program sources was performed in a 5-pixel radius aperture (corresponding to 0\farcs75 and 0\farcs69 radius software apertures at the Hale 5-meter and at the Keck II 10-meter telescopes, respectively). Aperture corrections were derived from the flux standards for each night, and applied to each target object's instrumental magnitude before application of the zero-point calibration and airmass correction. In the case of close companions, the combined system flux was first determined from a software aperture chosen to be large enough to contain both objects. Subsequently, the relative photometry of each component was determined by fitting a known point-source calibrator (generally one of the flux calibrators observed that night) to the individual source peaks. The total flux was then divided amongst the components in the ratio determined by the relative point-source fitting photometry.

\section{Analysis and Results}

Five of 13 (38\% $\pm$ 17\%) of the $\rho$ Oph sources (sepn. range $\sim$ 1\arcsec -- 10\farcs5 or $\sim$ 125 -- 1500 AU) and 3/6 (50\% $\pm$ 29\%) of the Serpens sources (sepn. range $\sim$ 1\farcs5 -- 5\arcsec or $\sim$ 470 -- 1560 AU) {\it appear} to be binary or multiple in nature. Thus, among the sources surveyed, we find 8/19 (42\% $\pm$ 15\%) stars in the separation range 1\arcsec -- 10\farcs5 ($\sim$ 125 -- 1560 AU) to be members of binary or higher order systems. The quoted uncertainties represent the statistical $\sqrt{N}$ errors in the fraction of sources found to be binary or multiple. In Figures~\ref{figure1} --~\ref{figure8}, we present our near and mid-IR images of each multiple source. IRS 43 was not detected at $J$-band, and we do not as yet have 10 $\mu$m data for L1689 SNO2 and EC 82/EC 86. Because the spacing of our on-chip chop/nod pattern was close to the component separations for GY 23/GY 21 and EC 92/EC 95, we do not present 10 micron images of these systems. The photometry was, nevertheless, recovered from the raw data. Contour levels for each image are listed in the respective figure captions. EC 82 forms a possible wide binary with EC 86. However, given their separation of $\sim$ 2700 AU at our adopted distance to Serpens of 310 pc, these objects may not compose a true bound binary system. We therefore exclude this system from our calculated CSF. Each source is individually discussed in detail in the next section.

In Tables~\ref{table2} and~\ref{table3}, we present the $JHKL$ magnitudes, 10 $\mu$m fluxes (in Janskys), near-IR colors, and spectral indices for all surveyed sources. Spectral indices were calculated from a least squares fit to the 2.2 $\mu$m to 10 $\mu$m data. We do not as yet have MIRLIN $N$-band fluxes for IRS 63, IRS 67, VSSG 17, L1689 SNO2, ISO 159, EC 53, and EC 82/EC 86. However, mid-IR ISO fluxes at 6.7 $\mu$m and 14.3 $\mu$m have been published for the $\rho$ Oph single sources IRS 67 and VSSG 17 by \cite{bon01}. The spectral indices reported in Table~\ref{table2} for these two sources were calculated using the ISO mid-IR fluxes. If neither MIRLIN nor ISO fluxes were available for a given source, no spectral index was calculated.

Seven sources in Tables~\ref{table2} and~\ref{table3} were not detected at $J$-band. This was not likely due to faint sources being smeared out in these relatively long (90 second) exposures, since this effect was not observed in other equally long $J$-band images in which faint sources were detected. Upper limits for the $J$-band magnitudes for these sources were determined by adding artificial stars to the respective $J$-band images, and counting the number of sources recovered by DAOFIND. Artificial stars were added at random positions to each image in twenty separate half magnitude bins with each bin containing one hundred stars. The twenty bins covered a magnitude range from 15.0 to 25.0. The artificial stars were examined to ensure that they had a similar FWHM of the point-spread function as the sources detected in other $J$-band images. Aperture photometry was performed on all sources to confirm that the assigned magnitudes of the added sources agreed with those returned by PHOT. All photometry agreed to within 0.1 magnitudes. DAOFIND and PHOT were then run and the number of identified artificial sources within each half magnitude bin was tallied. This process was repeated 20 times. Our final $J$-band magnitude limits, good to within $\pm$ 0.25 magnitudes, are listed in Tables~\ref{table2} and~\ref{table3}.

In Table~\ref{table4}, we list the separation (in arcsec) and position angle (measured with respect to the brightest source at $K$-band; the southernmost source in each case) for the sources in our survey which were found to be multiple. Figures~\ref{figure9} and~\ref{figure10} show the SEDs for WL 1 and SVS 20. In each case, the SED of the northernmost component has been shifted down by 1.0 in the vertical axis for clarity. The $\rho$ Oph source L1689 SNO2 and the Serpens sources EC 82/EC 86 were also found to be binary, however no resolved 10 $\mu$m fluxes are currently available for these objects, therefore SEDs were not constructed. We can obtain estimates for the SED classes of these objects by examining their location in $JHK$ and $JHKL$ color-color diagrams. In this respect, EC 86 is found to have near-IR colors indicative of a reddened background or Class III source (PMS star with no circumstellar material; ALS87). By contrast, the colors of both components of L1689 SNO2 and EC 82 show infrared excesses suggestive of Class II (TTS surrounded by a circumstellar disk; ALS87) objects or earlier. 

\section{Discussion}

\subsection{Notes on the Multiple Sources}

IRS 43 (also known as YLW 15) was found to be a binary VLA source with 0\farcs6 separation (\cite{grc00}). IRS 43 is also part of a wide-binary system with GY 263 (see \cite{all02} and Table~\ref{table3}). We find IRS 43 to be multiple at 10 $\mu$m (0\farcs51 separation; P.A. = 332.7 degrees) but single in the near-IR. The 2MASS catalog position (detected at $H$ and $K$) for this object is $\alpha = 16^h27^m26.^s94, \delta = -24^\circ40\arcmin50\arcmin\arcmin$ (J2000). This 2MASS position is accurate to better than 0\farcs2, and it is located very near the VLA 2 component in the 3.6cm map of this system presented by Girart et al. (2000). Therefore all the near-IR measurements of this object are likely associated only with the south-east $N$-band component (see Figure~\ref{figure1}). We have imaged IRS 43 at 2 $\mu$m wavelength with the 10-m Keck II telescope under conditions of 0\farcs5 seeing but still only detected a single source. We estimate that the companion is likely to be $K > 12$ mag from this observation and similar ones of others (L. Prato 2002, private communication).

The brighter mid-IR source in IRS 43 corresponds to VLA 2 and the heavily-veiled, Class I near-IR source. VLA 1, the resolved thermal jet component, is at the position of the fainter mid-IR source to the northwest. Both VLA/mid-IR sources associated with IRS 43 are embedded in extended, faint near-IR nebulosity imaged with HST/NICMOS (\cite{all02}). The total 1.3mm flux from IRS 43 is 75 mJy in an 11\arcsec \hspace*{0.05in}beam (\cite{am94}). This level of millimeter dust continuum emission is typical of Class I objects in $\rho$ Ophiuchus, and is therefore most likely associated with the heavily-veiled NIR/bright mid-IR source positionally coincident with VLA 2. Any protostar at an even earlier evolutionary stage than IRS 43/VLA 2 would have to have an order of magnitude greater millimeter flux than is observed (\cite{awb93}; \cite{awb00}). Given the strength of its mid-infrared and sub-millimeter emission, if an embedded, late-stage protostar were present in VLA 1, we would have detected it in the near-infrared images as well. The fact that we did not, argues for an outflow origin for the mid-infrared emission associated with VLA 1.

IRS 43 is, in fact, associated with a compact molecular outflow which is nearly pole-on (\cite{bon96}). The blue lobe is to the north-west, and the red lobe is displaced south-east. IRS 43 is also a hard X-ray source with strong flares (\cite{gro97}), and its total near-to far-IR bolometric luminosity is approximately 10 L$_{\odot}$ (\cite{wly89}). Greene \& Lada (2000, 2002) did not detect any photospheric absorption features in the high-resolution near-IR spectrum of this object, indicating a very large near-IR veiling which is consistent with a high mass accretion rate (\.{M} = 1.7 $\times$ 10$^{-6}$ M$_{\odot}$ yr$^{-1}$).

IRS 51 appears to be another example of a flat-spectrum source, in which the extended, possibly binary morphologies observed in the near-infrared and in the mid-infrared, do not agree with each other (see Figure~\ref{figure2}). Although IRS 51 is resolved along an approximately north-south direction at both wavelengths, the secondary component at $N$-band is at a different position angle (PA = 15$^\circ$) and separation ($\sim$ 0\farcs7) from the primary than the low-level extension observed towards the North in the $K$-band contours (PA = 10$^\circ$, separation $\sim$ 1\farcs5). This behaviour suggests, analogously to the case of IRS 43, that the mid-infrared ``secondary'' may very well be a bright knot in the molecular outflow from IRS 51 (\cite{bon96}), along a North-South cavity seen in scattered light at $K$-band. 

The near-to-far IR bolometric luminosity of IRS 51 is measured to be only 1.4 L$_{\odot}$. If this luminosity is entirely due to a single pre-main-sequence star without accretion, then PMS models indicate that its mass would be approximately 0.5 M$_{\odot}$. Greene \& Lada (2000) detected weak and broad 2.2935 $\mu$m CO band head absorption in the brighter $K$-band component. This is indicative of significant $K$-band veiling ($r_k > 1$) and fast rotation ($v$ sin $i \sim 50$ km s$^{-1}$). Luhman \& Rieke (1999) determined that the spectral type of this object is K5 -- K7, consistent with the low mass suggested from the total system bolometric luminosity.

The WL 1 system has the SED of a flat spectrum object. It was first resolved as a binary system in the near-IR by Ressler (1992), and reported as a mid-IR binary by Barsony \& Ressler (2000), who classified WL 1N as a Class II object and WL 1S as a flat spectrum object, based on the individual components' spatially-resolved 2-10 micron spectral indices. The WL 1 system has a bolometric luminosity of 1.8 L$_{\odot}$ (\cite{wly89}). It has a peak 1.3 mm flux of 25 mJy but is not associated with a molecular outflow (\cite{am94}). Luhman \& Rieke (1999) determined that the spectral type (of the brighter near-IR component) is later than M3, consistent with the low bolometric luminosity for the system.

The L1689 SN02 system is located in the L1689 dark cloud, southeast of the main L1688 $\rho$ Oph cloud and its embedded stellar cluster. This system is identified as L1689 \#5 by Greene et al. (1994) who determine that it has a flat 2--10 $\mu$m spectral index ($\alpha = 0.0$) and a 1.25--20 $\mu$m calorimetric luminosity lower limit of 1.3 L$_{\odot}$. Bontemps et al. (2001) estimate that its total stellar luminosity is 1.9 L$_{\odot}$ and its disk luminosity is 1.4 L$_{\odot}$. This source is approximately 2.5\arcmin \hspace*{0.05in}west of the peak DCO+ core position in the southern region of the L1689 cloud.

Eiroa et al. (1987) first found that the Serpens SVS 20 system was binary via CCD imaging at 9140\AA \hspace*{0.05in}and had extended emission in CCD images. These authors reported a 1\farcs7 separation at P.A. = 10$^\circ$. The binary was subsequently imaged by Huard, Weintraub, \& Kastner (1997). These authors found flux ratios of SVS 20(S)/SVS 20(N) = 2.89 +/-0.05, 2.99 +/- 0.03, and 2.87 +/- 0.03 at 3.3, 3.4, and 4.0 microns, respectively. For comparison, our corresponding 10.8 micron flux ratio is 2.85. Casali \& Eiroa (1996) found that its 2 $\mu$m $\delta v = 2$ CO bands were in emission weakly. The 2--14 $\mu$m ($K$ band and ISO) spectral index of SVS 20 is found to be flat ($\alpha = 0.0$) by Kaas (2001; private communication). Hurt \& Barsony (1996) found that the far-IR energy distribution of this object is also relatively flat over the 12--100 $\mu$m IRAS bands. They estimate that this system has 0.3 M$_{\odot}$ of circumstellar material. Casali et al. (1993) note that this source has weak sub-mm emission, detected at 800 and 1100 $\mu$m.  They estimate that the bolometric luminosity of this system is 0.6 -- 30 L$_{\odot}$. By integrating under the spatially resolved SEDs of the SVS 20 system presented here for the first time (see Table~\ref{table3} \& Figure~\ref{figure10}), we find L $=$ 4.1 L$_{\odot}$ for SVS 20N and L $=$ 12.5 L$_{\odot}$ for SVS 20S. These luminosity determinations represent only lower limits for the true bolometric luminosities of these sources, however, since the integrations were carried out over the limited wavelength range (1.25 -- 10.8 $\mu$m) over which we have spatially resolved data.

EC 92 and EC 95 in Serpens form a binary system at 5\farcs03 separation at P.A. $\simeq$ 352$^\circ$. EC 95 is by far the brighter at 10.8 micron, and is an 
extremely X-ray luminous proto-Herbig Ae/Be star (\cite{prei99}). 

GY 21 and GY 23 (sep'n = 10\farcs5; P.A. $\simeq$ 323$^\circ$) form a candidate wide binary system. GY 23 (also known as Elias 23) has been found to be variable by Elias (1978). GY 21 is associated with the centimeter source LFAM 3 (\cite{leo91}), is classified as a Class II M0 source by Wilking et al. (2001).

Finally, EC 82 (also known as SVS 2) forms a possible wide binary with EC 86 (sepn. = 8\farcs65; P.A. $\simeq$ 63$^\circ$), and displays extended nebulosity at $J$, $H$, and $K$-band. EC 82 is the illuminating source of the famous Serpens Reflection Nebula, and is especially apparent in our $J$ and $H$-band images (see Figure~\ref{figure8}).

\subsection{Multiplicity and SED Characteristics}

We define the companion star fraction (CSF) to be the ratio of the number of detected binary and multiple systems, to the observed number of single, binary, and multiple systems. For reasons discussed in the previous section, we consider IRS 51 to be a single source plus reflection nebulosity associated with the blue-shifted outflow lobe from the primary rather than a true binary star. Thus, we find a CSF for the ClassI/flat-spectrum sources surveyed of 7/19 (37\% $\pm$ 14\%). This is consistent with the CSFs derived for PMS stars in each of the Taurus, $\rho$ Ophiuchus, Chamaeleon, Lupus, and Corona Australis star-forming regions over a separation range of $\sim$ 150 -- 1800 AU (\cite{lei93}, \cite{simon95}, \cite{gmpb97}, \cite{all02}). In contrast, the CSF for solar-type main-sequence stars in the solar neighborhood in this separation range (11\% $\pm$ 3\%; \cite{dm91}, \cite{fm92}) is approximately one-third that of our sample. As is thought to be the case with TTSs, it is possible that this difference is due to evolutionary effects or environmental differences. For example, $\rho$ Ophiuchus and Serpens are dark cloud complexes, however some stars form in giant molecular cloud environments (e.g. \cite{lada92}, \cite{css95}, \cite{pl97}). Indeed, the observed binary fraction in Orion is consistent with that for solar-type main sequence stars (\cite{pro94}, \cite{petr98}, \cite{duch99}).

Finally, the composite SEDs of SVS 20 and WL 1 are indicative of Class I and flat spectrum sources, respectively. The individual components of SVS 20 both show flat spectrum SEDs, while both components of WL 1 appear to be Class II objects. Such behavior is consistent with what one typically finds for TTSs, where the companion of a classical TTS also tends to be a classical TTS (\cite{ps97}).

\section{Summary and Conclusions}

We have obtained new near- and mid-IR observations of 19 Class I/flat-spectrum objects in the $\rho$ Ophiuchi and Serpens dark clouds. These observations are part of a larger systematic infrared multiplicity survey of Class I/flat-spectrum objects in the nearest dark clouds. The primary results and conclusions derived from the present study can be summarized as follows:

1. The CSF for the Class I/flat-spectrum sources surveyed is 7/19 (37\% $\pm$ 14\%). These results are consistent with the CSFs derived for PMS stars in each of the Taurus, $\rho$ Ophiuchus, Chamaeleon, Lupus, and Corona Australis star-forming regions over a similar separation range. However, the CSF for solar-type main-sequence stars in the solar neighborhood in this separation range (11\% $\pm$ 3\%) is approximately one-third that of our sample, and may be attributed to evolutionary effects or environmental differences.

2. Among the sources which were found to be binary, and for which SEDs of both sources could be constructed, we find that the individual components of SVS 20 and WL 1 exhibit the same SED classifications. This behaviour is similar to what one typically finds for TTSs, where the companion of a classical TTS also tends to be a classical TTS (\cite{ps97}).

3. The strength of the mid-infrared and sub-millimeter emission observed in IRS 43/VLA 1 suggests that, if an embedded, late-stage protostar were present in this source, we would have detected it in our near-infrared images as well; we did not. In addition, the near- and mid-IR ``secondary'' sources observed in IRS 51 are at different position angles and separations from each other. We therefore suggest that the companion sources observed in IRS 43 and IRS 51 may not be stellar objects, but rather are associated with molecular outflows from the primary YSO, although IRS 43 does form a wide binary with GY 263.

\acknowledgements

We wish to thank Amanda Kaas for providing near-IR data for Class I/flat-spectrum sources in Serpens in advance of publication. We also thank the IRTF staff for their outstanding support in making our observations possible. This research has made use of the NASA/IPAC Infrared Science Archive, which is operated by the Jet Propulsion Laboratory, California Institute of Technology, under contract with the National Aeronautics and Space Administration. We thank the Keck Observatory Director and staff for making it possible to use MIRLIN on the Keck II telescope. The staff of Palomar Observatory have continued to provide outstanding support for MIRLIN as a visiting instrument. K. E. H. gratefully acknowledges support from a National Research Council Research Associateship Award. M. B. gratefully acknowledges financial support through NSF grant AST-0096087 (CAREER), from NASA's Long Term Space Astrophysics Research Program, NAG5 8412, and the NASA/ASEE Summer Faculty Fellowship program at the Jet Propulsion Laboratory, which made her contributions to this work possible. T. P. G. acknowledges grant support from the NASA Origins of Solar Systems Program, NASA RTOP 344-37-22-11.

\newpage
\figcaption[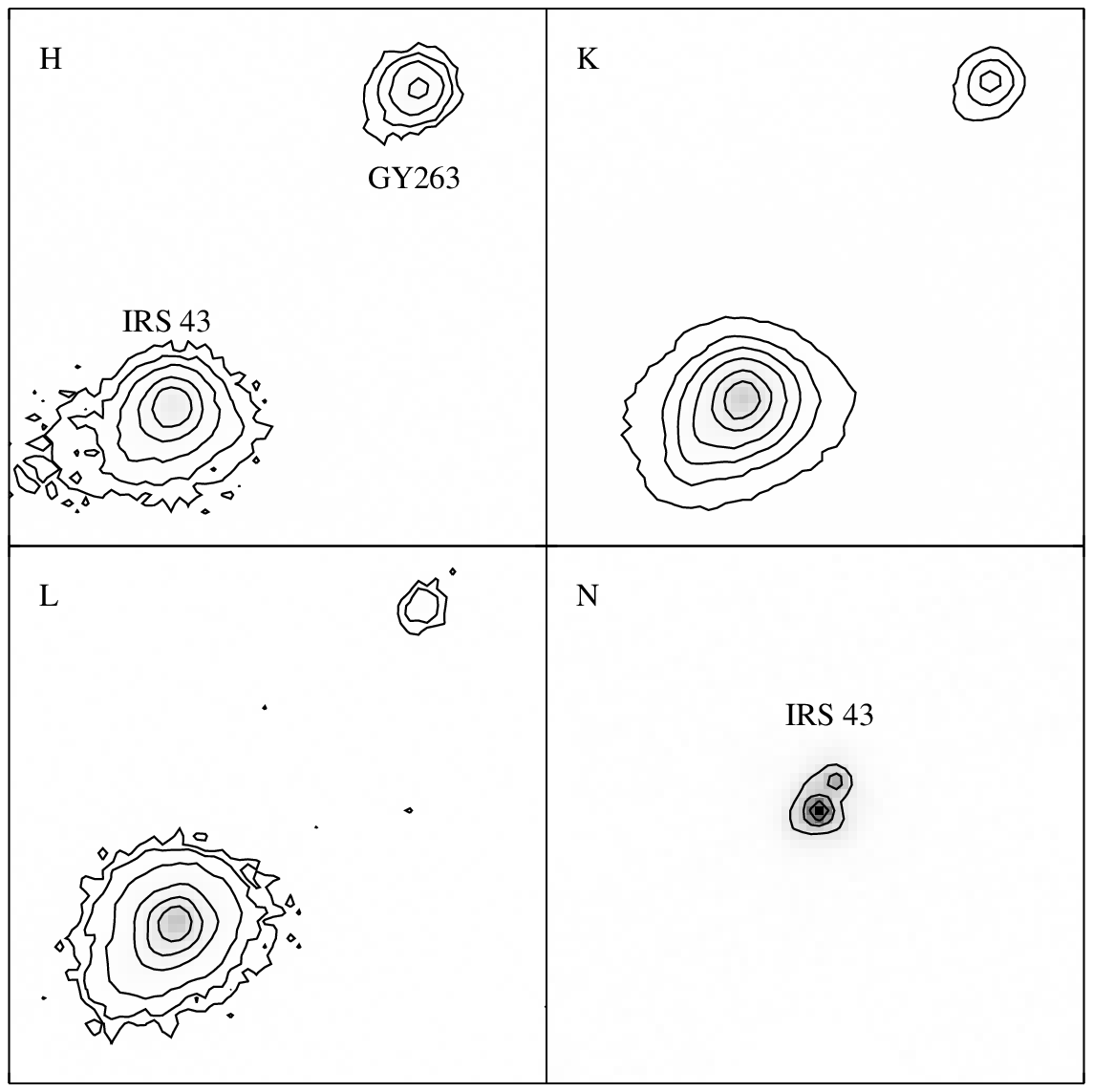]
{$H$, $K$, $L$, and $N$-band images of the field containing IRS 43. The wide ($\sim$ 7\farcs0 separation; see Table~\ref{table3}) binary with GY 263 is apparent in the $H$, $K$, and $L$-bands. Neither component of this wide binary was detected at $J$-band (see Table~\ref{table2}). North is up, and East is to the left in each image. Pixel sizes are 0\farcs148 at $HKL$ and 0\farcs138 at $N$-band, for a total field of view at $HKL$ of $\sim$ 9\farcs5 $\times$ 9\farcs5, and 8\farcs7 $\times$ 8\farcs7 at $N$-band. Contour levels for the $H$-band image are 2.5, 5, 10, 25, and 50\% of the peak value; contour levels for the $K$-band image are 1, 2.5, 5, 10, 25, and 50\% of the peak value; contour levels in the $L$-band image are 0.5, 1, 2.5, 10, 25,and 50\% of the peak value. Contour levels in the $N$-band image are at 10, 25, and 50\% of the peak value. Our MIRLIN observations resolved two distinct components at 0\farcs51 separation at a position angle of 332.7 degrees in IRS 43. The southern component coincides positionally with VLA 2; the fainter, northern component coincides with VLA 1, the resolved thermal radio jet (\cite{grc00}). This component was not resolved in our near-IR images. \label{figure1}
}

\figcaption[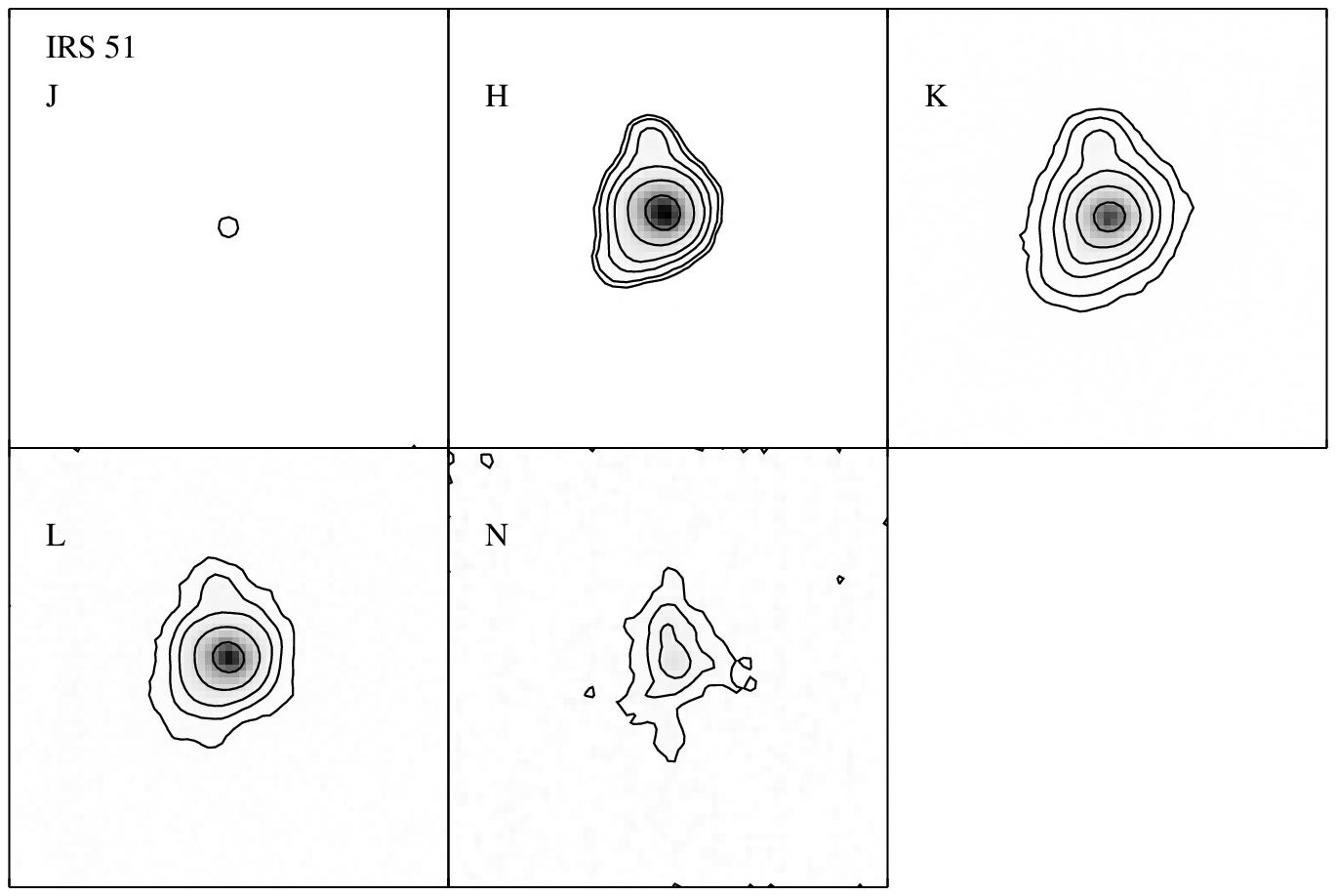]
{$J$, $H$, $K$, $L$, and $N$-band images of IRS 51. The field of view of each image is approximately 9\farcs5 $\times$ 9\farcs5. North is up, and East is to the left in each image. Contour levels are 25, 50, and 75\% of the peak value in the $J$-band image; 0.5, 1, 2.5, 5, 15, and 50\% of the peak value in the $H$ and $K$-band images; 1, 2.5, 5, 15, and 50\% of the peak in the $L$-band image; and 10, 25, and 50\% of the peak value in the $N$-band image. Although IRS 51 is resolved along an approximately north-south direction at both near- and mid-IR wavelengths, the secondary component at $N$-band is at a different position angle (PA = 15$^\circ$) and separation ($\sim$ 0\farcs7) from the primary than the low-level extension observed towards the North in the $H$, $K$, and $L$-band contours (PA = 10$^\circ$, separation $\sim$ 1\farcs5).
\label{figure2}
}

\figcaption[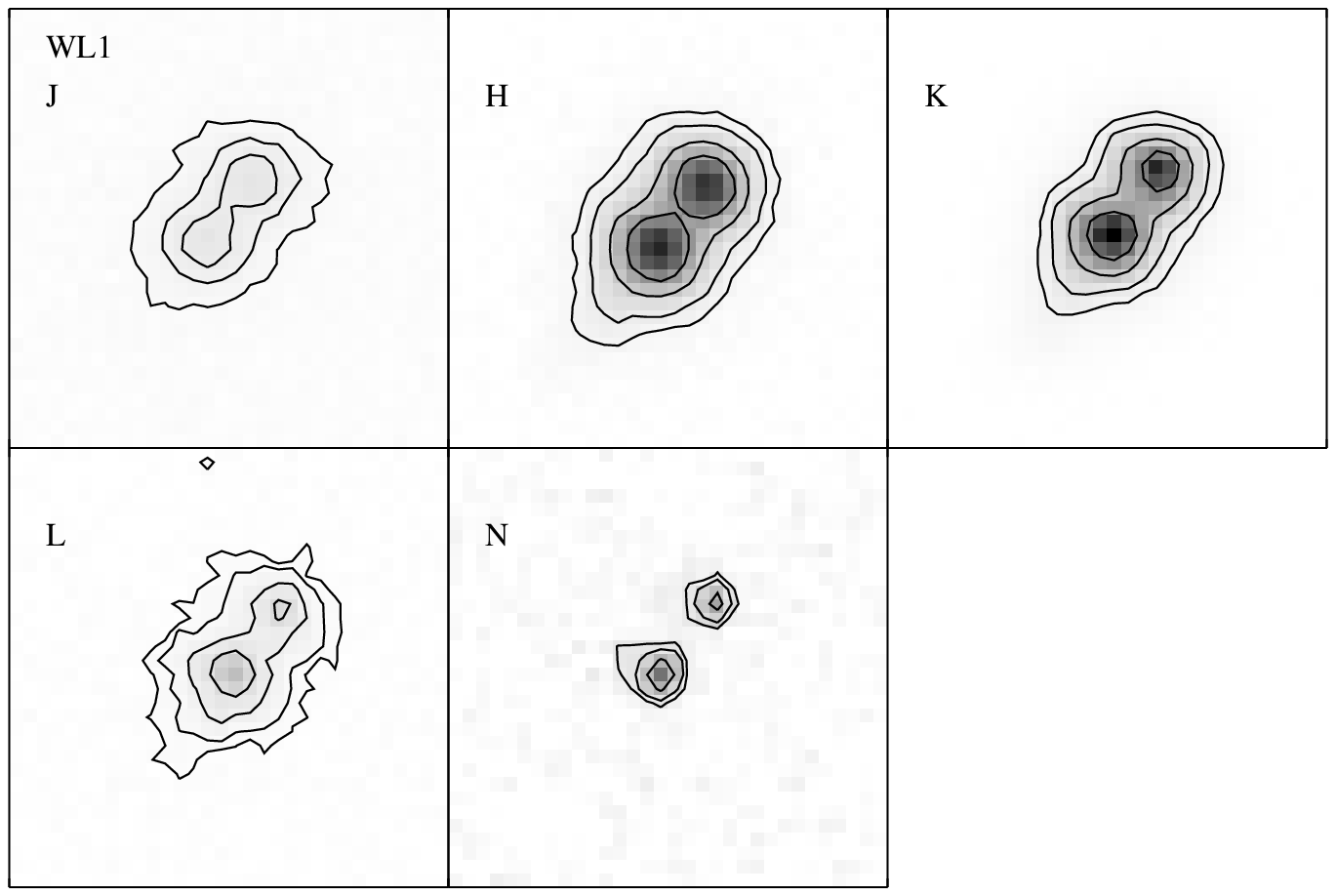]
{$J$, $H$, $K$, $L$, and $N$-band images of the WL 1 binary. The field of view of the $JHKL$ images is approximately 4\farcs6 on a side with 0\farcs148 pixels, while the $N$-band image is 4\farcs3 on a side with 0\farcs138 pixels. North is up, and East is to the left in each image. Contour levels in the $J$-band image are at 25, 50, and 75\% of the peak value; contour levels in the $H$, $K$, $L$-band images are at 5, 10, 25, and 50\% of the respective peak values, and the $N$-band image contour levels are at 15, 25, and 50\% of the peak value. Note that although both sources are about equal brightness at $J$-band, the southern source gradually increases in brightness towards the longer wavelengths, even as the northern source progressively dims towards the longer wavelengths. While the combined SED for WL 1 implies a flat-spectrum object \cite{bon01}, the individual components of this source exhibit pronouncedly Class II SEDs.
\label{figure3}
}

\figcaption[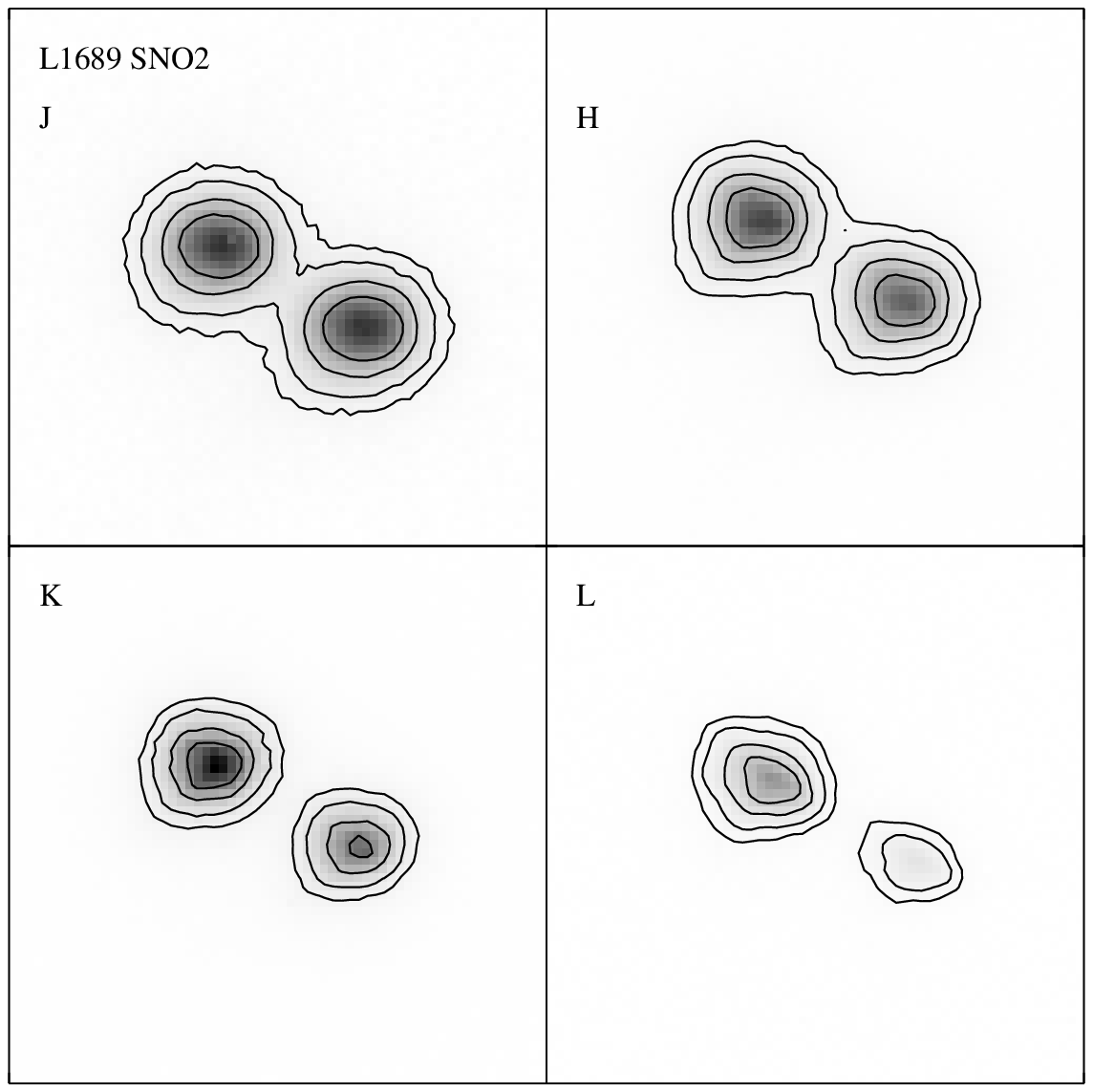]
{$J$, $H$, $K$, and $L$-band images of L1689 SNO2. Each image is approximately 9\farcs5 on a side. North is up, and East is to the left in each image. Contour levels are 5, 10, 25, and 50\% of the peak value in all panels. Although we do not as yet have resolved 10-micron imaging photometry, note the differing behavior of the sources with wavelength: although both are nearly equally bright at $J$-band, the southern source dims much faster towards longer wavelengths than its northern neighbor.
\label{figure4}
}

\figcaption[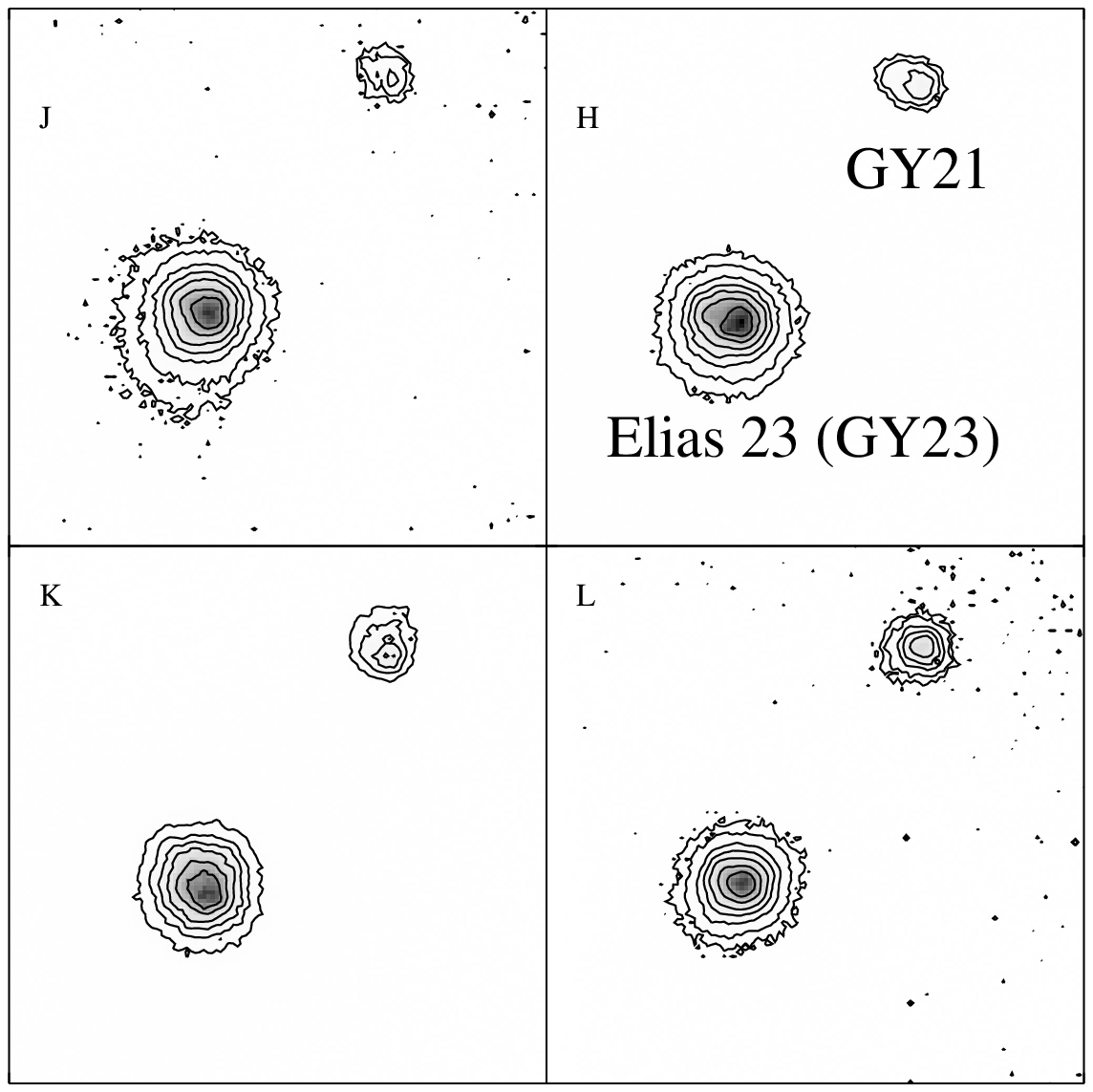]
{$J$, $H$, $K$, and $L$-band images of the wide binary GY21/GY23. Each image is approximately 19\arcsec on a side. North is up, and East is to the left in each image. Contour levels for the $JH$ and $K$-band images are 0.5, 1, 2.5, 5, 10, 25, and 50\% of the peak value. Contour levels in the $L$-band image are at 1, 2.5, 5, 10, 25, and 50\% of the peak value.
\label{figure5}
}

\figcaption[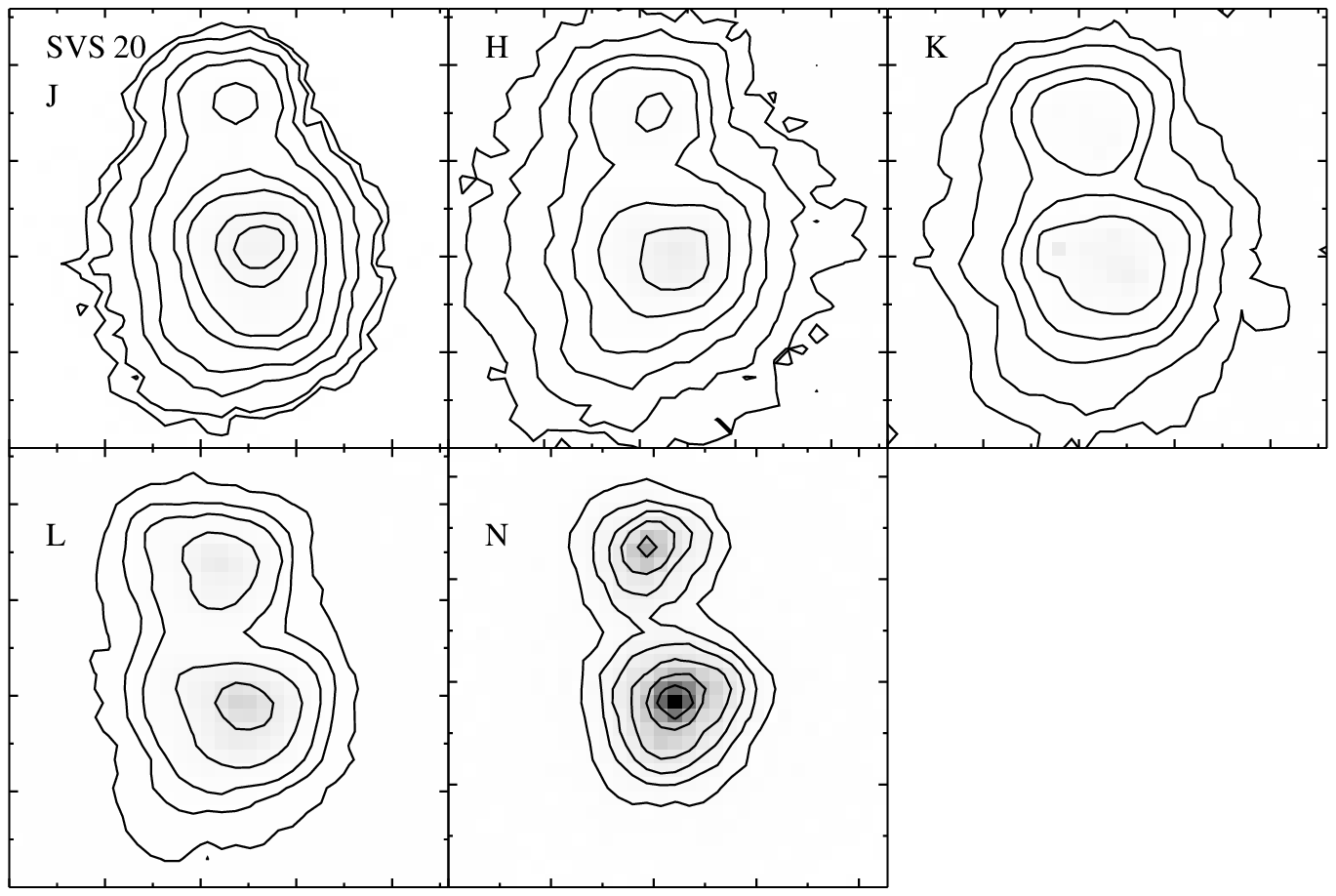]
{$J$, $H$, $K$, $L$, and $N$-band images of SVS 20. The pixel scale of the $JHKL$ images is 0\farcs148 pixel$^{-1}$, and 0\farcs138 pixel $^{-1}$ for the $N$-band image. Each image is therefore approximately 4\farcs7 on a side. North is up, and East is to the left in each image. Contour levels in the $J$-band image are at 0.5, 1, 2.5, 5, 12, 20, 50, and 75\% of the peak value; contour levels in the $H$ and $K$-band images are at 0.5, 1, 2.5, 5, 15, and 50\% of the respective peak values. The $K$-band image was smoothed with a 3 $\times$ 3 boxcar. Contour levels in the $L$-band image are at 1, 2.5, 5, 15, and 50\% of the peak value, and the $N$-band image contour levels are at 1, 2.5, 5, 10, 25, and 50\% of the peak value. The individual components of SVS20 are separated by 1\farcs51, at a position angle of 9.9 degrees.
\label{figure6}
}

\figcaption[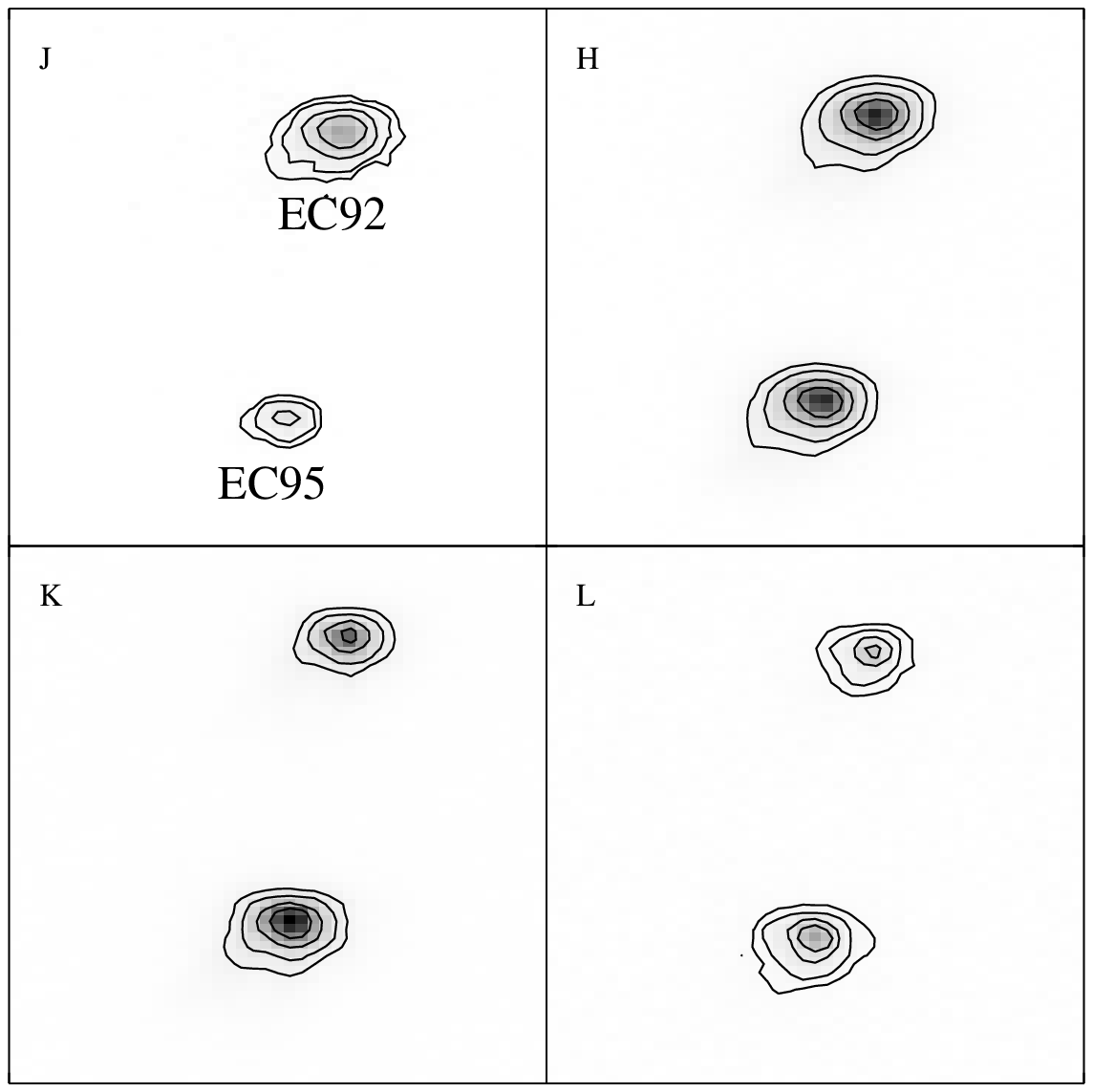]
{$J$, $H$, $K$, and $L$-band images of the wide binary EC92/EC95. Each image is approximately 6\farcs7 $\times$ 9\farcs5. North is up, and East is to the left in each image. Contour levels are 5, 10, 25, and 50\% of the peak value in all panels.
\label{figure7}
}

\figcaption[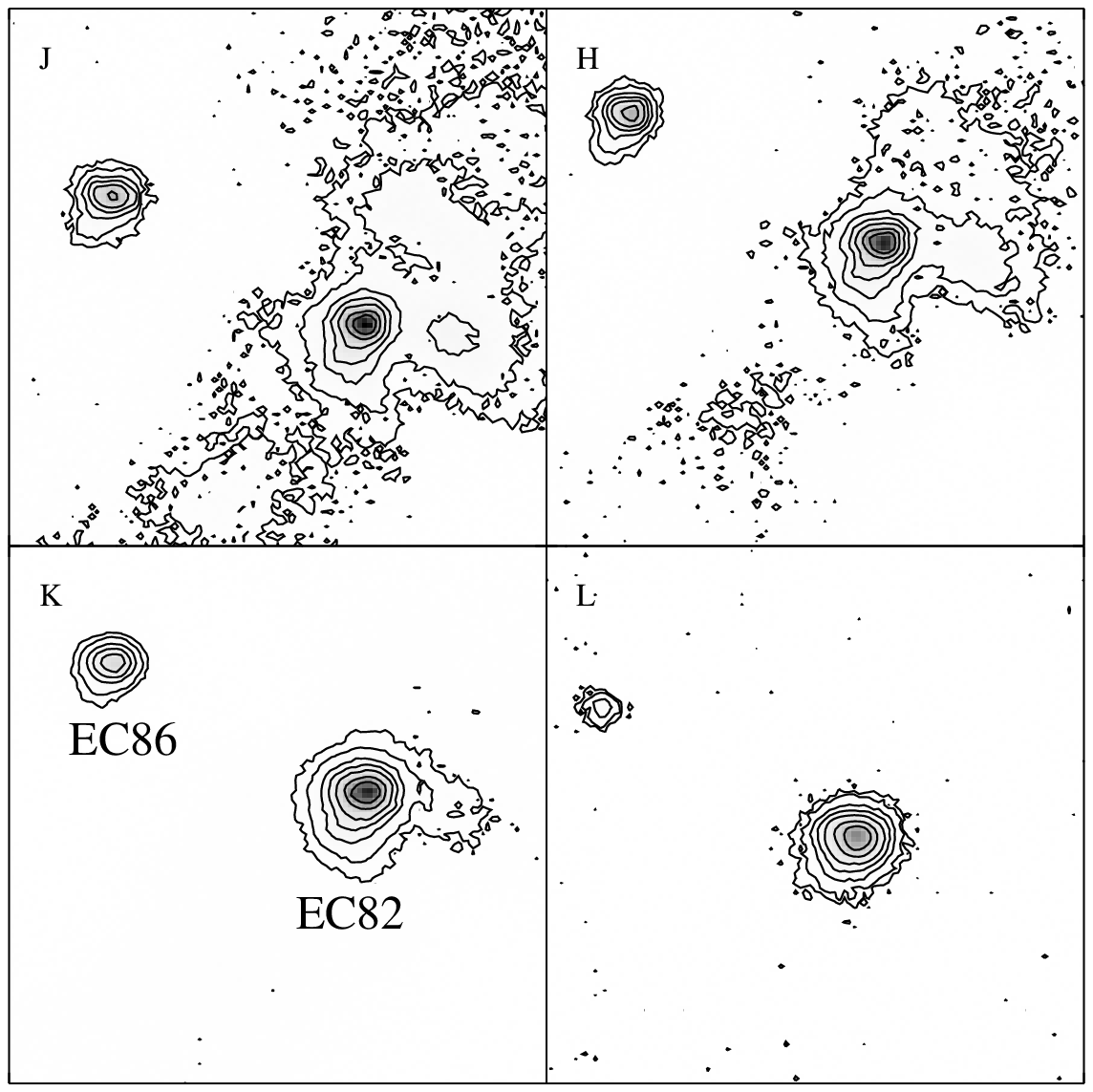]
{$J$, $H$, $K$, and $L$-band images of the possible wide binary EC82/EC86. Each image is approximately 16\farcs3 on a side. North is up, and East is to the left in each image. Contour levels are 0.5, 1, 2.5, 5, 10, 25, and 50\% of the peak value in all panels. The famous Serpens Reflection Nebula is especially apparent in the $J$ and $H$-band images.
\label{figure8}
}

\figcaption[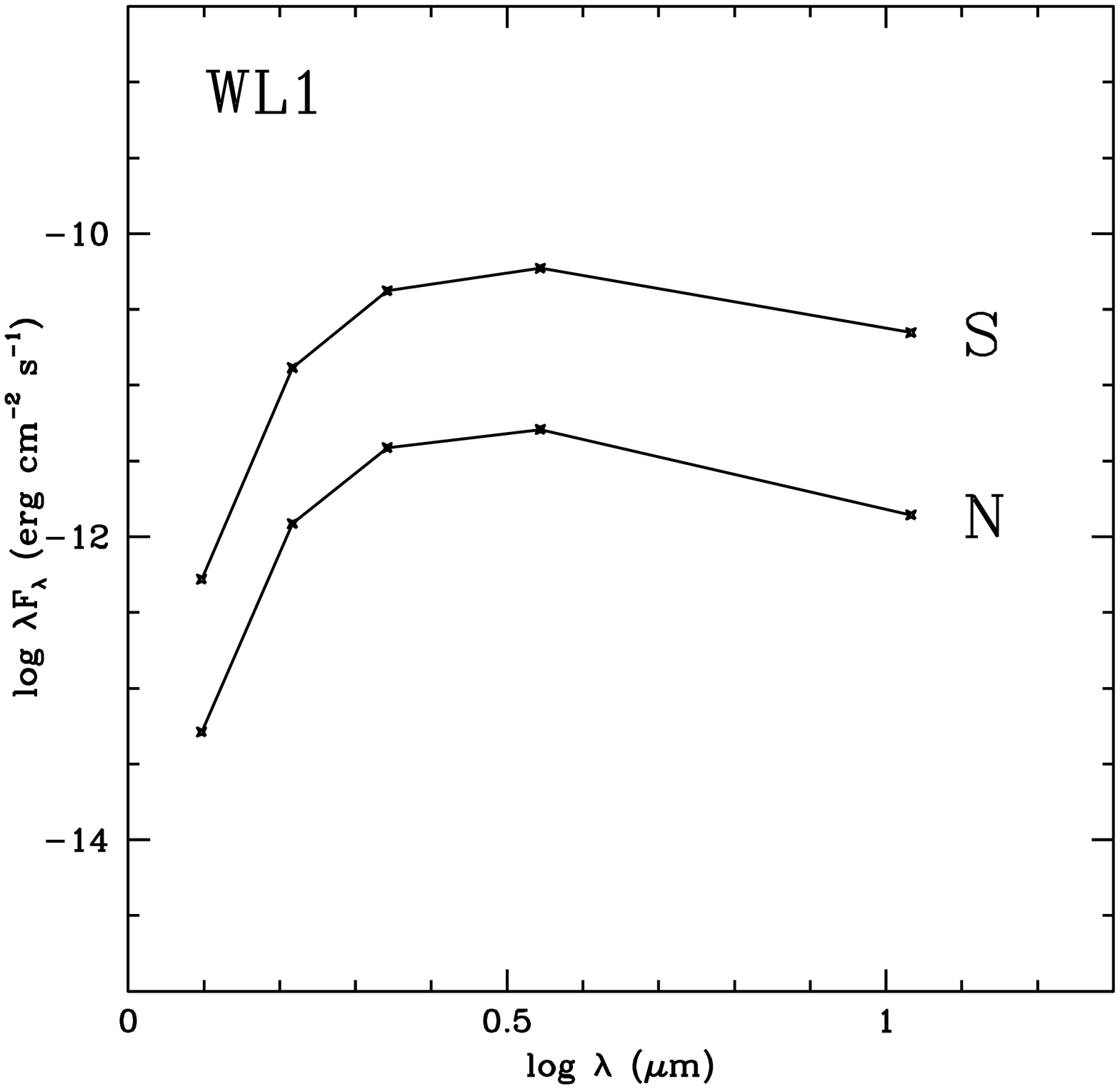]
{Spectral Energy Distribution of the $\rho$ Ophiuchus protostellar binary WL 1. The SED of the northern component (N) has been shifted down by 1.0 in the vertical axis for clarity. Both components of this source exhibit Class II SEDs.
\label{figure9}
}

\figcaption[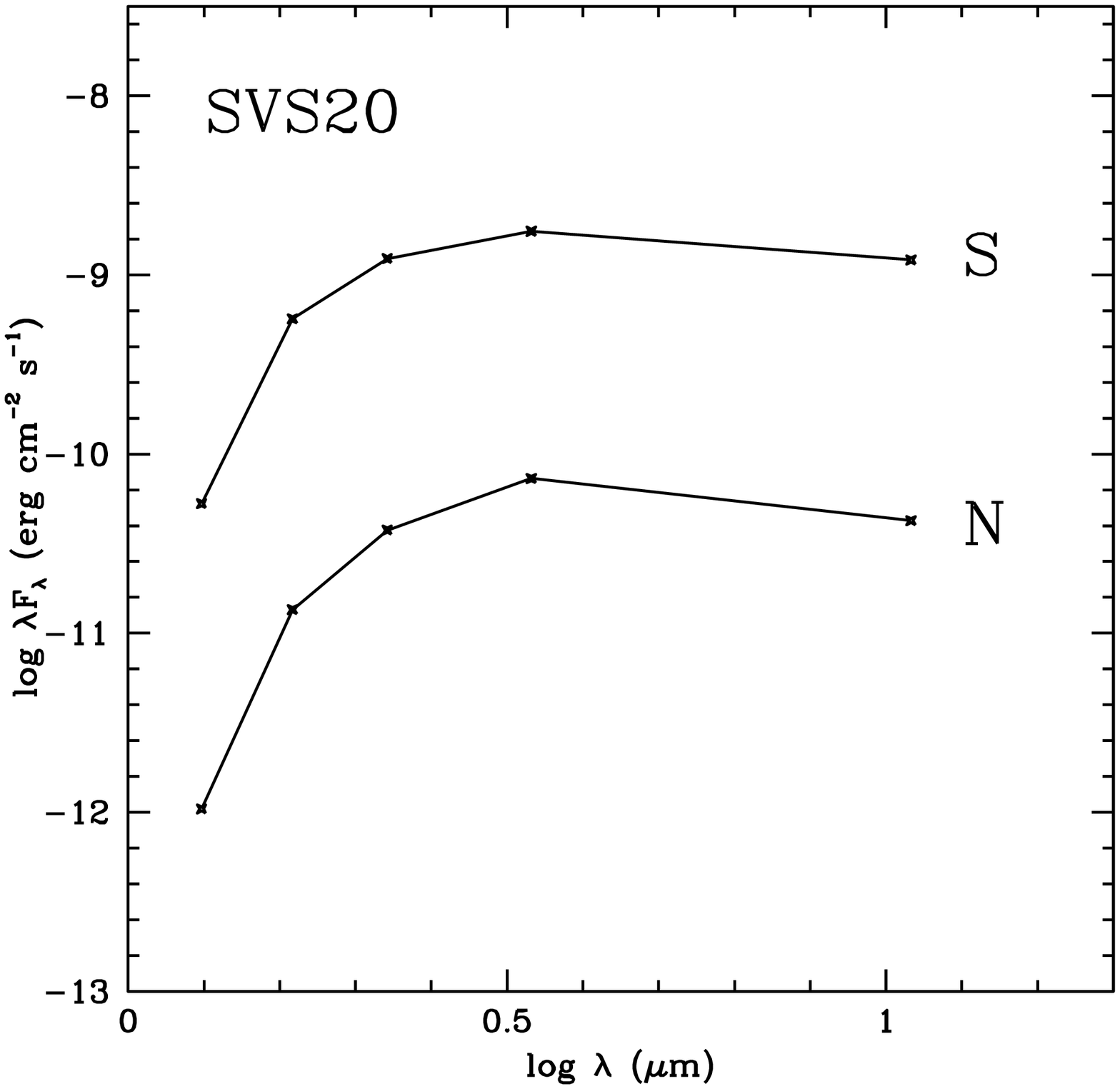]
{Spectral Energy Distribution of the Serpens protostellar binary SVS 20. The SED of the northern component (N) has been shifted down by 1.0 in the vertical axis for clarity. Both components of this source exhibit flat spectrum SEDs.
\label{figure10}
}

\clearpage
\begin{deluxetable}{lccccc}
\footnotesize
\tablecaption{Summary of Observations \label{table1}}
\tablewidth{0pt}
\tablehead{Source & Region & RA (2000) & Dec (2000) & IRTF $JHKL$ UT Date & $N$-band UT Date/Location}
\startdata
IRS 43 & $\rho$ Oph & 16 27 26.90 & -24 40 51.50 & 2001 Jul 09 & 1998 Jun 07/Keck II\nl
 & & & & & \nl
GY 263 & $\rho$ Oph & 16 27 26.60 & -24 40 45.90 & 2001 Jul 09 & 1998 Jun 07/Keck II\nl
 & & & & & \nl
IRS 67 & $\rho$ Oph & 16 32 01.00 & -24 56 44.00 & 2001 Jul 09 & Not Observed\nl
 & & & & & \nl
IRS 51 & $\rho$ Oph & 16 27 39.84 & -24 43 16.10 & 2001 Jul 09 & 1997 Jun 25/Palomar\nl 
 & & & & & \nl
YLW 16A & $\rho$ Oph & 16 27 28.00 & -24 39 34.30 & 2001 Jul 10 & 1997 Jun 26/Palomar\nl
 & & & & & \nl
GY 224 & $\rho$ Oph & 16 27 11.17 & -24 40 46.70 & 2001 Jul 10 & 1998 Jun 07/Keck II\nl
 & & & & & \nl
IRS 63 & $\rho$ Oph & 16 31 35.53 & -24 01 28.30 & 2001 Jul 10 & Not Observed\nl
 & & & & & \nl
WL 1 & $\rho$ Oph & 16 27 04.13 & -24 28 30.70 & 2001 Jul 10 & 1998 Jun 07/Keck II\nl 
 & & & & & \nl
WL 19 & $\rho$ Oph & 16 27 11.74 & -24 38 32.10 & 2001 Jul 11 & 1996 Apr 23/Palomar\nl
 & & & & & \nl
VSSG 17 & $\rho$ Oph & 16 27 30.18 & -24 27 44.30 & 2001 Jul 11 & Not Observed\nl
 & & & & & \nl
GY 21 & $\rho$ Oph & 16 26 23.54 & -24 24 41.50 & 2001 Jul 11 & 1998 Jul 02/Palomar\nl
 & & & & & \nl
GY 23 & $\rho$ Oph & 16 26 24.00 & -24 24 49.86 & 2001 Jul 11 & 1998 Jul 02/Palomar\nl
 & & & & & \nl
L1689SNO2 & $\rho$ Oph & 16 31 52.13 & -24 56 15.20 & 2001 Jul 11 & Not Observed\nl
 & & & & & \nl
EC 92 & Serpens & 18 29 57.75 & 01 12 56.96 & 2001 Jul 09 & 1998 Jun 07/Keck II\nl
 & & & & & \nl
EC 95 & Serpens & 18 29 57.80 & 01 12 52.00 & 2001 Jul 09 & 1998 Jun 07/Keck II\nl
 & & & & & \nl
ISO 159 & Serpens & 18 29 32.00 & 01 18 42.00 & 2001 Jul 09 & Not Observed\nl
 & & & & & \nl
SVS 20 & Serpens & 18 29 57.70 & 01 14 07.00 & 2001 Jul 09 & 1998 Jun 07/Keck II\nl 
 & & & & & \nl
EC 53 & Serpens & 18 29 51.20 & 01 16 42.00 & 2001 Jul 10 & Not Observed\nl
 & & & & & \nl
EC 82 & Serpens & 18 29 56.80 & 01 14 46.00 & 2001 Jul 10 & Not Observed\nl
\enddata
\end{deluxetable}

\clearpage
\begin{deluxetable}{lrrrrcrccr}
\footnotesize
\tablecaption{$JHKL$ Magnitudes, 10$\mu$m flux, Near-IR Colors, and Spectral Indices for $\rho$ Oph Sources \label{table2}}
\tablewidth{0pt}
\tablehead{Source & J & H & K & L & $F_{10}$$_{\mu}$$_{m}$\tablenotemark{a} & (J-H) & (H-K) & (K-L) &
$\alpha$\tablenotemark{b}}
\startdata
IRS 43 S & $>$ 19.26 & 12.56 &  9.44 &  6.77 & 1.54 & $>$ 6.70 & 3.12 & 2.67 & 0.6\nl
IRS 43 N & ---- & ---- & ---- & ---- & 0.52 & --- & --- & --- & ---\nl
 & & & & & & & & & \nl
GY 263 & $>$ 19.29 & 14.19 & 12.42 & 10.71 & 0.03 & $>$ 5.10 & 1.77 & 1.71 & -0.2\nl
 & & & & & & & & & \nl
IRS 67 & $>$ 20.26 & 14.06 & 10.34 &  7.80 & ---- & $>$ 6.20 & 3.72 & 2.54 & 0.7\tablenotemark{c}\nl
 & & & & & & & & & \nl
IRS 51 S & 17.60 & 11.04 &  8.69 &  6.88 & 0.73 & 6.56 & 2.35 & 1.81 & -0.3\nl 
IRS 51 N & $>$ 19.77 & 12.65 & 11.18 &  9.17 & 0.48 & $>$ 7.12 & 1.47 & 2.01 & 1.0\nl
 & & & & & & & & & \nl
YLW 16A & $>$ 19.27 & 14.72 & 10.48 & 7.77 & 2.65 & $>$ 4.55 & 4.24 & 2.71 & 1.6\nl
 & & & & & & & & & \nl
GY 224 & 20.04 & 13.33 & 10.84 &  8.51 & 0.25 & 6.71 & 2.49 & 2.33 & 0.3\nl
 & & & & & & & & & \nl
IRS 63 & 16.73 & 11.76 &  9.34 &  7.18 & ---- & 4.97 & 2.42 & 2.16 & ---\nl
 & & & & & & & & & \nl
WL 1 S & 17.21 & 12.84 & 10.76 & 9.02 & 0.08 & 4.37 & 2.08 & 1.74 & -0.5\nl 
WL 1 N & 17.23 & 12.91 & 10.85 & 9.18 & 0.05 & 4.32 & 2.06 & 1.67 & -0.7\nl
 & & & & & & & & & \nl
WL 19 & $>$ 19.28 & 15.02 & 11.18 & 8.32 & 0.18 & $>$ 4.26 & 3.84 & 2.86 & 0.2\nl
 & & & & & & & & & \nl
VSSG 17 & 17.05 & 11.34 & 9.07 & 6.85 & ---- & 5.71 & 2.27 & 2.22 & 0.3\tablenotemark{c}\nl
 & & & & & & & & & \nl
GY 21 & 14.59 & 11.87 & 10.09 & 8.04 & 0.40 & 2.72 & 1.78 & 2.05 & 0.1\nl
GY 23 & 10.74 & 8.59 & 7.39 & 5.92 & 2.02 & 2.15 & 1.20 & 1.47 & -0.4\nl
 & & & & & & & & & \nl
L1689 SNO2 S & 12.13 & 10.05 & 8.86 & 7.96 & ---- & 2.08 & 1.19 & 0.90 & ---\nl
L1689 SNO2 N & 12.16 & 9.81 & 8.32 & 6.64 & ---- & 2.35 & 1.49 & 1.68 & ---\nl 
\enddata
\tablenotetext{a}{10 $\mu$m fluxes are listed in Janskys.}
\tablenotetext{b}{Spectral index not calculated if both MIRLIN and ISO (\cite{bon01}) fluxes unavailable.}
\tablenotetext{c}{Spectral index calculated using the 6.7 $\mu$m and 14.3 $\mu$m ISO fluxes from \cite{bon01}.}
\end{deluxetable}

\clearpage
\begin{deluxetable}{lrrrrcrccr}
\footnotesize
\tablecaption{$JHKL$ Magnitudes, 10$\mu$m flux, Near-IR Colors, and Spectral Indices for Serpens Sources \label{table3}}
\tablewidth{0pt}
\tablehead{Source & J & H & K & L & $F_{10}$$_{\mu}$$_{m}$\tablenotemark{a} & (J-H) & (H-K) & (K-L) &
$\alpha$\tablenotemark{b}}
\startdata
EC 92 & 14.71 & 12.03 & 10.44 & 8.96 & 0.55 & 2.68 & 1.59 & 1.48 & 0.7\nl
EC 95 & 15.93 & 11.90 & 9.78 & 8.34 & 0.10 & 4.03 & 2.12 & 1.44 & -0.4\nl
 & & & & & & & & & \nl
ISO 159 & 11.25 & 9.30 & 8.33 & 6.68 & ---- & 1.95 & 0.97 & 1.65 & ---\nl
 & & & & & & & & & \nl
SVS 20 S & 12.20 & 8.74 & 7.09 & 5.34 & 4.36 & 3.46 & 1.65 & 1.75 & -0.1\nl 
SVS 20 N & 13.96 & 10.30 & 8.38 & 6.29 & 1.53 & 3.66 & 1.92 & 2.09 & 0.0\nl
 & & & & & & & & & \nl
EC 53 & $>$ 19.31 & 16.32 & 14.65 & 11.30 & ---- & $>$ 2.99 & 1.67 & 3.35 & ---\nl
 & & & & & & & & & \nl
EC 82 & 11.75 & 10.32 & 8.99 & 7.24 & ---- & 1.43 & 1.33 & 1.75 & ---\nl
EC 86 & 12.85 & 11.57 & 10.90 & 10.49 & ---- & 1.95 & 0.67 & 0.41 & ----\nl
\enddata
\tablenotetext{a}{10 $\mu$m fluxes are listed in Janskys.}
\tablenotetext{b}{No spectral index calculated if 10 $\mu$m flux unavailable.}
\end{deluxetable}

\clearpage
\begin{deluxetable}{lcc}
\footnotesize
\tablecaption{Separations and Position Angles for the Binary Sources \label{table4}}
\tablewidth{0pt}
\tablehead{Source & Separation (\arcsec) & Position Angle
(degrees)\tablenotemark{a}}
\startdata
IRS 43/GY 263 & 6.99 & 322.0\nl 
 & & \nl
WL 1 & 0.82 & 321.2\nl 
 & & \nl
GY 23/GY 21 & 10.47 & 322.6\nl 
 & & \nl
L1689 SNO2 & 2.92 & 240.3\nl 
 & & \nl
SVS 20 & 1.51 & 9.9\nl
 & & \nl
EC 95/EC 92 & 5.03 & 352.1\nl 
\enddata
\tablenotetext{a}{Measured with respect to the brightest source at $K$-band; the
southernmost source in all cases.}
\end{deluxetable}

\clearpage

\clearpage
\plotone{Haisch.fig1.ps}
\clearpage
\plotone{Haisch.fig2.ps}
\clearpage
\plotone{Haisch.fig3.ps}
\clearpage
\plotone{Haisch.fig4.ps}
\clearpage
\plotone{Haisch.fig5.ps}
\clearpage
\plotone{Haisch.fig6.ps}
\clearpage
\plotone{Haisch.fig7.ps}
\clearpage
\plotone{Haisch.fig8.ps}
\clearpage
\plotone{Haisch.fig9.eps}
\clearpage
\plotone{Haisch.fig10.eps}
\clearpage


\newpage
\begin{thebibliography}{}

\bibitem[Abt \& Levy 1976]{al76} Abt, H. A. \& Levy, S. 1976, \apjs, 30, 273
\bibitem[Adams, Lada, \& Shu 1988]{als88} Adams, F. C., Lada, C. J., \& Shu, F. H. 1988, \apj, 326, 865. 
\bibitem[Adams, Lada, \& Shu 1987]{als87} Adams, F. C., Lada, C. J., \& Shu, F. H. 1987, \apj, 321, 788
\bibitem[Allen et al.\ 2002]{all02} Allen, L. E., Myers, P. C., Di Francesco, J., Mathieu, R., Chen, H., \& Young, E. 2002, \apj, 566, 993
\bibitem[Andr\'{e} \& Montmerle 1994]{am94} Andr\'{e}, P. \& Montmerle, T. 1994, \apj, 420, 837
\bibitem[Andr\'{e}, Ward-Thompson, \& Barsony 2000]{awb00} Andre, P., Ward-Thompson, D., \& Barsony, M. 2000, in Protostars and Planets IV, 59
\bibitem[Andr\'{e}, Ward-Thompson, \& Barsony 1993]{awb93} Andre, P., Ward-Thompson, D., \& Barsony, M. 1993, \apj, 406, 122
\bibitem[Barsony, Ressler, \& Marsh 2002]{brm02} Barsony, M., Ressler, M. E., \& Marsh, K. 2002, in preparation
\bibitem[Barsony \& Ressler 2000]{br00} Barsony, M. \& Ressler, M. E.\ 2000, IAU Symposium, 200, 47P
\bibitem[Bontemps et al.\ 2001]{bon01} Bontemps, S., Andr\'{e}, P., Kaas, A. A., Nordh, L., Olofsson, G., Huldtgren, M., Abergel, A., Blommaert, J., Boulanger, F., Burgdorf, M., Cesarsky, C. J., Cesarsky, D., Copet, E., Davies, J., Falgarone, E., Lagache, G., Montmerle, T., P\'{ e}rault, M., Persi, P., Prusti, T., Puget, J. L., \& Sibille, F. 2001, \aap, 372, 173
\bibitem[Bontemps et al.\ 1996]{bon96} Bontemps, S., Andr\'{e}, P., Terebey, S., \& Cabrit, S. 1996, \aap, 311, 858
\bibitem[Brandner \& Zinnecker 1997]{bz97} Brandner, W. \& Zinnecker, H. 1997, \aap, 321, 220
\bibitem[Carpenter, Snell \& Schloerb 1995]{css95} Carpenter, J. M., Snell, R. L., \& Schloerb, F. P.  1995, \apj, 450, 201
\bibitem[Casali \& Eiroa 1996]{ce96} Casali, M. M. \& Eiroa, C. 1996, \aap, 306, 427
\bibitem[Casali, Eiroa, \& Duncan 1993]{ced93} Casali, M. M., Eiroa, E., \& Duncan, W. D. 1993, \aap, 275, 195
\bibitem[de Lara, Chavarria-K., \& L\'{o}pez-Molina 1991]{del91} de Lara, E., Chavarria-K. C., \& Lopez-Molina, G. 1991, \aap, 243, 139
\bibitem[Duch{\^ e}ne 1999 ]{duch99} Duch{\^ e}ne, G. 1999, \aap, 341, 547
\bibitem[Duquennoy \& Mayor 1991]{dm91} Duquennoy, A., \& Mayor, M. 1991, \aap, 248, 484
\bibitem[Eiroa et al.\ 1987]{eir87} Eiroa, C., Lenzen, R., Leinert, Ch., \& Hodapp, K. W. 1987, \aap, 179, 171
\bibitem[Elias et al.\ 1982]{el82} Elias, J. H., Frogel, J. A., Matthews, K., \& Neugebauer, G. 1982, \aj, 87, 102
\bibitem[Elias 1978]{el78} Elias, J. H. 1978, \apj, 224, 453
\bibitem[Fischer \& Marcy 1992]{fm92} Fischer, D. A., \& Marcy, G. W. 1992, \apj, 396, 798
\bibitem[Ghez et al.\ 1997]{gmpb97} Ghez, A. M., McCarthy, D. W., Patience, J., \& Beck, T. 1997, \apj, 481, 378
\bibitem[Ghez, Neugebauer, \& Matthews 1993]{gnm93} Ghez, A. M., Neugebauer, G., \& Matthews, K. 1993, \aj, 106, 2005
\bibitem[Girart, Rodr\'{i}guez, \& Curiel 2000]{grc00} Girart, J, M., Rodr\'{i}guez, L. F., \& Curiel, S. 2000, \apj, 544, L153
\bibitem[Greene \& Lada 2002]{gl02} Greene, T. P., \& Lada, C. J. 2002, \aj, submitted
\bibitem[Greene \& Lada 2000]{gl00} Greene, T. P., \& Lada, C. J. 2000, \aj, 120, 430
\bibitem[Greene et al.\ 1994]{greene94} Greene, T. P., Wilking, B. A., Andr\'{e}, P., Young, E. T., \& Lada, C. J. 1994, \apj, 434, 614
\bibitem[Grosso et al.\ 1997]{gro97} Grosso, N., Montmerle, T., Feigelson, E. D., Andre, P., Casanova, S., \& Gregorio-Hetem, J. 1997, \nat, 387, 56
\bibitem[Hartigan, Strom, \& Strom 1994]{hss94} Hartigan, P., Strom, K. M., \& Strom, S. E. 1994, \apj, 427, 961
\bibitem[Huard, Weintraub, \& Kastner 1997]{hua97} Huard, T. L., Weintraub, D. A., \& Kastner, J. H. 1997, \mnras, 290, 598
\bibitem[Hurt \& Barsony 1996]{hb96} Hurt, R. L. \& Barsony, M. 1996, \apj, 460, L45
\bibitem[Knude \& Hog 1998]{kh98} Knude, J. \& Hog, E. 1998, \aap, 338, 897.
\bibitem[Lada 1987]{lada87} Lada, C. J. 1987, in Star Forming Regions, ed. M. Peimbert \& J. Jugaku (Dordrecht: Reidel), 1
\bibitem[Lada 1992]{lada92} Lada, E. A. 1992, \apj, 393, L25
\bibitem[Leinert et al.\ 1993]{lei93} Leinert, C., Zinnecker, H., Weitzel, N., Christou, J., Ridgway, S. T., Jameson, R., Haas, M., \& Lenzen, R. 1993, \aap, 278, 129
\bibitem[Leous et al.\ 1991)]{leo91} Leous, J. A., Feigelson, E. D., Andr\'{e}, P., \& Montmerle, T. 1991, \apj, 379, 683
\bibitem[Looney, Mundy, \& Welch 2000]{lmw00} Looney, L. W., Mundy, L. G., \& Welch, W. J. 2000, \apj, 529, 477
\bibitem[Looney, Mundy, \& Welch 1997]{lmw97} Looney, L. W., Mundy, L. G., \& Welch, W. J. 1997, \apj, 484, L157
\bibitem[Luhman \& Rieke 1999]{lr99} Luhman, K. L. \& Rieke, G. H. 1999, \apj, 525, 440
\bibitem[Mathieu 1994]{math94} Mathieu, R. D. 1994, \araa, 32, 465
\bibitem[Petr et al.\ 1998]{petr98} Petr, M. G., Coude Du Foresto, V., Beckwith, S. V. W., Richichi, A., \& McCaughrean, M. J. 1998, \apj, 500, 825
\bibitem[Phelps \& Lada 1997]{pl97} Phelps, R. L., \& Lada, E. A. 1997, \apj, 477, 176
\bibitem[Prato \& Simon 1997]{ps97} Prato, L. \& Simon, M. 1997, \apj, 474, 455
\bibitem[Preibisch 1999]{prei99} Preibisch, T. 1999, \aap, 345, 583
\bibitem[Prosser et al.\ 1994]{pro94} Prosser, C. F., Stauffer, J. R., Hartmann, L., Soderblom, D. R., Jones, B. F., Werner, M. W., \& McCaughrean, M. J. 1994, \apj, 421, 517
\bibitem[Rayner et al.\ 1993]{ray93} Rayner, J. T., Shure, M. A., Toomey, D. W., Onaka, P. M., Denault, A. J., Stahlberger, W. E., Watanabe, D., Criez, K., Robertson, L., Cook, D., \& Kidger, M. J. 1993, \procspie, 1946, 490
\bibitem[Ressler et al.\ 1994]{ress94} Ressler, M. E., Werner, M. W., Van Clever, J., \& Chou, H. A. 1994, Exp. Astr., 3, 277
\bibitem[Ressler 1992]{ress92} Ressler, M. E. 1992, Ph. D. Thesis, University of Hawaii
\bibitem[Ressler \& Barsony 2001]{rb01} Ressler, M. E. \& Barsony, M. 2001, \aj, 121, 1098
\bibitem[Shure et al.\ 1994]{shu94} Shure, M., Toomey, D. W., Rayner, J., Onaka, P., Denault, A., Stahlberger, W., Watanabe, D., Criez, K., Robertson, L., \& Cook, D. 1994, Exp. Astr., 3, 239
\bibitem[Simon et al.\ 1995]{simon95} Simon, M., Ghez, A. M., Leinert, Ch., Cassar, L., Chen, W. P., Howell, R. R., Jameson, R. F., Matthews, K., Neugebauer, G., \& Richichi, A. 1995, \apj, 443, 625
\bibitem[Wilking et al.\ 2001]{wilk01} Wilking, B. A., Bontemps, S., Schuler, R. E., Greene, T. P., \& Andr{\' e}, P. 2001, \apj, 551, 357
\bibitem[Wilking, Lada, \& Young 1989]{wly89} Wilking, B. A., Lada, C. J., \& Young, E. T. 1989, \apj, 340, 823
\bibitem[Zinnecker, McCaughrean, \& Wilking 1993]{zmw93} Zinnecker, H., McCaughrean, M. J., \& Wilking, B. A. 1993, in Protostars and Planets III, ed. E. H. Levy \& J. I. Lunine (Tucson: Univ. Arizona Press), 429

\end{thebibliography}
\end{document}